\documentclass[letterpaper, 10 pt, conference]{ieeeconf}
\IEEEoverridecommandlockouts
\usepackage{graphicx} % for pdf, bitmapped graphics files
\usepackage{amsmath} % assumes amsmath package installed
\usepackage{amssymb}  % assumes amsmath package installed
\usepackage{mathtools}

\usepackage{xcolor}
\DeclareMathOperator*{\argmax}{arg\,max}

\usepackage[hidelinks]{hyperref}
\newtheorem{thm}{Theorem}[section]

\newtheorem{lem}[thm]{Lemma}
\newtheorem{prop}[thm]{Proposition}
\newtheorem{rmk}{Remark}[section]
\newtheorem{dfn}{Definition}[section]
\newtheorem{asmp}{Assumption}[section]

\renewcommand{\QED}{\QEDopen}

\graphicspath{{./figures/}}

\begin{document}

\title{\LARGE \bf
	Data-Driven Fault Isolation in Linear Time-Invariant Systems: \\ 
	A Subspace Classification Approach
}

\author{Mohammad Amin Sheikhi, Gabriel de Albuquerque Gleizer, Peyman Mohajerin Esfahani and Tamás Keviczky% <-this % stops a space
	\thanks{This work was supported by the Digital Twin project with project number P18-03 of the research programme TTW-Perspective which is (partly) financed by the Dutch Research Council (NWO).}% <-this % stops a space
	\thanks{$^{1}$The authors are with the Department of Delft Center for Systems and Control, Delft University of Technology, Mekelweg 2, 2628 CD Delft, The Netherlands \{\tt\small m.a.sheikhi,\, g.gleizer,\,p.mohajerinesfahani, and t.keviczky\}{\tt\small @tudelft.nl}\,\scriptsize.}%
}

\maketitle
\thispagestyle{empty}

%%%%%%%%%%%%%%%%%%%%%%%%%%%%%%%%%%%%%%%%%%%%%%%%%%%%%%%%%%%%%%%%%%%%%%%%%%%%%%%%
\begin{abstract}
We study the problem of fault isolation in linear systems with actuator and sensor faults within a data-driven framework. We propose a nullspace-based filter that uses solely fault-free input-output data collected under process and measurement noises. By reparameterizing the problem within a behavioral framework, we achieve a direct fault isolation filter design that is independent of any explicit system model. The underlying classification problem is approached from a geometric perspective, enabling a characterization of mutual fault discernibility in terms of fundamental system properties given a noise-free setting. In addition, the provided conditions can be evaluated using only the available data. Finally, a simulation study is conducted to demonstrate the effectiveness of the proposed method.
\end{abstract}

%\begin{IEEEkeywords}
%	Data-driven control, Fault diagnosis, Linear systems, Subspace methods.
%\end{IEEEkeywords}

%%%%%%%%%%%%%%%%%%%%%%%%%%%%%%%%%%%%%%%%%%%%%%%%%%%%%%%%%%%%%%%%%%%%%%%%%%%%%%%%
\section{INTRODUCTION}
Fault detection and isolation are essential components of modern health monitoring systems, driven by the ever-increasing demand for improved reliability and safety in complex industrial systems. Fault detection is the task of determining whether the system is healthy or experiencing a fault during the real operation. On top of it, \emph{fault isolation} (FI) -- the focus of this letter -- is the more involved task of identifying the root cause of the faulty behavior.

To assess the state of operation, diagnosis filters should ignore the influence of all external inputs on the system output, except for the fault. The resulting signal is called \emph{residual} \cite{varga2023solving}. Both model-based and data-driven approaches for residual generation have been developed in the literature. Recently, the latter has received most attention, since in practice large datasets are available, but accurate mathematical models are scarce and costly. Classical data-driven methods rely on system identification followed by model-based diagnosis filter design. More recently, \emph{behavioral} methods based on on Willems' Fundamental Lemma~\cite{willems1986fundamentallemma} enable a direct filter design, bypassing the identification step \cite{markovsky2021behavioral}. %
This approach has recently been used in \cite{disaro2024UIOequivalence,fattore2024data} to develop a data-driven \textit{unknown input observer}.% -- a residual generation filter obtained by solving an input-decoupling problem -- for fault diagnosis. 
Another important family of diagnosis filters is based on nullspace (or parity-space) \cite{chow1984parityspace, nyberg2006nullspaceDAE}, \cite{krishnan2020data}, which we build upon in this letter. These methods exploit the left nullspace of the extended observability matrix to filter out the contribution of the system internal states to the output. Nullspace methods enable building a bank of filters for FI, see, e.g., \cite{esfahani2015probabilisticDAE, ding2014ddSKR, shang2022DistRobustFDSKR}, by ensuring that each filter is sensitive to only one specific fault rather than all faults \cite{ding2009subspace}. 

While the theoretical foundations of nullspace-based FI filters have been extensively studied (e.g., \cite{varga2023solving,nyberg2006nullspaceDAE}, \cite{massoumnia2002failure}), their  performance limits in a data-driven setting remain largely unexplored; more importantly, fault discernbility is not fully understood. Additionally, state-of-the-art bank-of-filters approaches, which rely on scalar residual signals, may impose limitations on filter performance \cite{ding2014ddSKR}. This letter aims to bridge these gaps within a data-driven framework.

\textbf{Contributions}: We employ a geometric approach based on behavioral nullspaces to design a FI scheme for sensor and actuator faults. Moreover, we characterize fault discernibility under a noise-free condition and show that fault signals corresponding to transmission zeros of certain subsystems are indiscernible, causing isolation ambiguities. Notably, the indiscernible fault subspaces can be verified using only data.

\textbf{Notation}. Throughout this letter, $\mathbb{R}$, $\mathbb{C}$, and $\mathbb{N}_0$ denote the sets of real, complex, and natural numbers, respectively. The nullspace of a matrix is represented by $\mathcal{N}(\cdot)$, while $\mathcal{R}(\cdot)$ denotes its range. The inner product of two vectors in the Euclidean space is given by $\langle \cdot, \cdot\rangle$, and the Euclidean-norm is denoted by $\lVert \cdot \rVert$. The Kronecker product is represented by $\otimes$, and the symbol $I_n$ shows the identity matrix of size $n$.

\section{NULLSPACE BASED RESIDUAL GENERATION}
\subsection{System}
A general discrete-time finite-dimensional linear time-invariant (LTI) system, in the presence of additive faults, can be represented in the following innovation form~\cite{verhaegen1994innovationform}
\begin{equation}\label{eq:general-LTI-ss}
	\begin{aligned}
		x_{k+1} &= Ax_k+B_uu_k+B_ff_k+Ke_k\,, \\
		y_k &=Cx_k+D_uu_k+D_ff_k+e_k\,.
	\end{aligned}
\end{equation}
where the tuple $(A,[\,B_u,\,B_f],\,C,\,[D_u,\,D_f])$ represents the minimal state-space realization of the system matrices with appropriate dimensions. The state vector $x_k$ is $n$-dimensional, corresponding to the order of the system. The input, output, and fault signals are denoted by $u_k\in\mathbb{R}^{n_u}$, $y_k\in\mathbb{R}^{n_y}$, and $f_k\in\mathbb{R}^{n_f}$, respectively. $K\in \mathbb{R}^{n\times n_y}$ is a steady-state Kalman gain, and $e_k\in \mathbb{R}^{n_y}$ is the zero-mean innovation process with covariance matrix $\Sigma_e$. The $i$th fault component can be modeled as either a sensor or an actuator fault as follows:
\begin{equation}\label{eq:fault-matrices-def}
\begin{split}
		&j\text{th actuator fault: } B^{[i]}_f =  B^{[j]}_u\,, \quad D^{[i]}_f = D^{[j]}_u \,,\\
		&j\text{th sensor fault: } B^{[i]}_f = 0\,,\quad D^{[i]}_f =I^{[j]}_{n_y}\,,
	\end{split}
\end{equation}
where superscripts $[i]$ and $[j]$ refer to the $i$th and $j$th columns of the associated matrices. Next, we define the required definitions and assumptions for our method.
\begin{asmp}\label{asmp:system-observability}
	The pair $(A,\,C)$ corresponding to the system~\eqref{eq:general-LTI-ss} is observable.
\end{asmp}
\begin{dfn}[Left-invertibility]\cite{kirtikar2011delayleftinvert}
    A system defined by an $n_y\times n_u$ proper transfer function $\mathbf{G}(\mathrm{z}):=C(\mathrm{z}I-A)^{-1}B_u+D_u$ is \textit{$\tau$-delay left invertible} if there exists an $n_u\times n_y$ proper transfer function $\mathbf{G}^{+}_{\tau}(\mathrm{z})$ such that $\mathbf{G}^{+}_{\tau}(\mathrm{z})\mathbf{G}(\mathrm{z})=\mathrm{\mathrm{z}}^{-\tau}I_{n_u}$ for almost all $\mathrm{z}\in\mathbb{C}$ and nonnegative integer $\tau$. If one such $\tau$ exists, the system is simply called \emph{left invertible}.
\end{dfn}
A necessary condition for left invertibility is $n_y\geq n_u$.
\begin{asmp}\label{asmp:leftinvert-faultsubsys}
    The fault subsystem $(A,\, B_f,\, C,\, D_f)$ is $\tau$-delay left invertible.
\end{asmp}

Assumption~\ref{asmp:leftinvert-faultsubsys} implies that $n_y\geq n_f\geq 2$, as the problem concerns \emph{fault isolation}.
\begin{prop}\cite[Theorem~1]{kirtikar2011delayleftinvert}
    The following statements are equivalent:
    {
    \renewcommand{\labelenumi}{(\roman{enumi})}
	\begin{enumerate}
		\item The system $(A,B_u,C,D_u)$ is left invertible.
		\item $\mathrm{rank}\begin{bmatrix}
			A-zI  &B_u \\
			C &D_u
		\end{bmatrix}=n+n_u$ for almost all $z\in\mathbb{C}$.
	\end{enumerate}
	}
\end{prop}

Let the notation $\mathbf{w}_{k_1,k_2}=[\begin{matrix}
	w^\top_{k_1} &\cdots &w^\top_{k_1+k_2-1}
\end{matrix}]^\top$ describe a time window of data points associated with an arbitrary signal $w_k$. The measurements over a sliding window of length $L$ are given by the following data equation
\begin{equation}\label{eq:dataequation-overwindow}
	\mathbf{y}_{k,L} = \mathcal{O}_Lx_k + \mathcal{T}^u_L\mathbf{u}_{k,L} +  \mathcal{T}^f_L\mathbf{f}_{k,L} + \mathcal{T}^e_L\mathbf{e}_{k,L},
\end{equation} 
in which $\mathcal{O}_L$ represents the extended observability and $\mathcal{T}^\star_L$ is the lower triangular block-Toeplitz matrix structured as
\begin{equation*}
	\begin{split}
		&\mathcal{O}_L=\begin{bmatrix}
			C \\ CA \\ \vdots \\ CA^{L-1}
		\end{bmatrix}\,, \quad \mathcal{T}^\star_L = \begin{bmatrix}
			M^\star_0 &0 &\cdots &0 \\
			M^\star_1 &M^\star_0 &\ddots &\vdots \\
			\vdots &\vdots &\ddots &0 \\
			M^\star_{L-1} &M^\star_{L-2} &\cdots &M^\star_0
		\end{bmatrix}\,
	\end{split}
\end{equation*}
with $\star$ representing $u$, $f$, or $e$. The corresponding Markov parameters are given by
\begin{equation*}\label{eq:MPs-definition}
	\begin{split}
		&M^u_i=\begin{cases}
			D_u & i=0\\
			CA^{i-1} B_u & i>0
		\end{cases}\,,\; M^f_i=\begin{cases}
			D_f & i=0\\
			CA^{i-1}B_f& i>0
		\end{cases}\,, \\
		&M^e_i=\begin{cases}
			I_{n_y} & i=0\\
			CA^{i-1} K & i>0
		\end{cases}\,.
	\end{split}
\end{equation*}
Let $N$ represent the total number of samples in the dataset. Denote by $\mathbf{W}_{k,L}=[\begin{matrix} \mathbf{w}_{k,L} &\mathbf{w}_{k+1,L} &\cdots &\mathbf{w}_{k+N-L,L}  \end{matrix}]$ the corresponding Hankel matrix of an arbitrary signal $w_k$, and $X_{k_1,k_2}=[\begin{matrix} x_{k_1} &x_{k_1+1} &\cdots &x_{k_1+N-k_2} \end{matrix}]$, consisting of system states. The Hankel format of the data equation~\eqref{eq:dataequation-overwindow} is obtained by
\begin{equation}\label{eq:dataequation-hankel-overwindow}
	\mathbf{Y}_{k,L} = \mathcal{O}_LX_{k,L} + \mathcal{T}^u_L\mathbf{U}_{k,L} +  \mathcal{T}^f_L\mathbf{F}_{k,L} + \mathcal{T}^e_L\mathbf{E}_{k,L}\,,
\end{equation}
leading to
\begin{align}\label{eq:dataequation-hankel-augmented}
	\begin{bmatrix}
		\mathbf{U}_{k,L} \\
		\mathbf{Y}_{k,L}
	\end{bmatrix} =\mathcal{G}_L\begin{bmatrix}
		\mathbf{U}_{k,L} \\
		X_{k,L}
	\end{bmatrix}+\begin{bmatrix}
		0 \\ \mathcal{T}^f_L\mathbf{F}_{k,L} + \mathcal{T}^e_L\mathbf{E}_{k,L}
	\end{bmatrix}\,,
\end{align}
with $\mathcal{G}_L:= \begin{bmatrix}
	I &0\\ \mathcal{T}^u_L &\mathcal{O}_L
\end{bmatrix},$ providing a compact representation of the underlying fault/noise-free system.

\subsection{Residual}
Nullspace-based diagnosis filter design \cite{ding2014ddSKR} requires identifying the left nullspace of $\mathcal{G}_L$ from the given fault-free input-output (I/O) data, $\{u_k,\,y_k\}^N_{k=1}$. This subspace is described by $\mathcal{K}_L$, i.e., $\mathcal{R}(\mathcal{K}^\top_L)=\mathcal{N}(\mathcal{G}^{\top}_L)$. Behavioral system theory gives that, for any healthy noise-free trajectory of the system, the \emph{kernel representation} is given by \cite{willems1997behavioraltheory}
\begin{equation}\label{eq:dd-KR}
	\forall \mathbf{u}_{k,L},\, x_k,\quad \mathcal{K}_{L}\begin{bmatrix}
		\mathbf{u}_{k,L} \\ \mathbf{y}_{k,L}
	\end{bmatrix}=0,\; \text{and}\; \mathcal{K}_{L}\begin{bmatrix}
	\mathbf{U}_{k,L} \\
	\mathbf{Y}_{k,L}
	\end{bmatrix} =0 \,.
\end{equation}

By partitioning $\mathcal{K}_L$ into $[\begin{matrix}\mathcal{K}^u_L &\mathcal{K}^y_L\end{matrix}]$ and using \eqref{eq:dataequation-hankel-augmented}, the following relationships hold:
\begin{equation*}
	\mathcal{K}^{y}_L\mathcal{O}_L=0;\quad \mathcal{K}^{u}_L=-\mathcal{K}^{y}_L\mathcal{T}^u_L\,,
\end{equation*}
using the fact that $\mathcal{K}_L\mathcal{G}_L=0$. $\mathcal{K}^{y}_L$ is referred to as the \emph{parity space} in the literature \cite{chow1984parityspace}. Rewriting \eqref{eq:dataequation-overwindow} into form of \eqref{eq:dataequation-hankel-augmented} for $[\mathbf{u}^{\top}_{k,L},\,\mathbf{y}^{\top}_{k,L}]^{\top}$, and pre-multiplying by $\mathcal{K}_L$ results in the residual signal
\begin{equation*}\label{eq:residual-signal}
	 r_k=\mathcal{K}_L\begin{bmatrix}
	 	\mathbf{u}_{k,L}\\ \mathbf{y}_{k,L}\end{bmatrix}=\mathcal{K}^{y}_L(\mathcal{T}^f_L\mathbf{f}_{k,L} + \mathcal{T}^e_L\mathbf{e}_{k,L})\,.
\end{equation*}
The detection filter, therefore, is entirely designed by obtaining a proper $\mathcal{K}_L$ satisfying \eqref{eq:dd-KR}. The subspace represented by $\mathcal{K}_L$ can be consistently estimated when the I/O data is corrupted by noise. For details on designing and optimizing such a detection filter, we refer readers to \cite{ding2014ddSKR,shang2022DistRobustFDSKR}. However, in this letter, we focus on the \emph{fault isolation} task, which follows immediately after the detection.

\section{FAULT ISOLATION SCHEME}

Given $\mathcal{K}_L$ satisfying \eqref{eq:dd-KR}, the resulting residual signal $r_k$ carries all the system's fault information. One key observation is that $r_k$ is possibly multi-dimensional, depending on the nullity of $\mathcal{G}^{\top}_L$, i.e., $\mathrm{dim}\,\mathcal{N}(\mathcal{G}^{\top}_L)$. In this sense, each row of $\mathcal{K}_L$ acts as a single-output finite-impulse-response (FIR) detection filter.. This fact has been utilized in designing bank of filters in \cite{ding2009subspace,esfahani2015probabilisticDAE}, which can discard information. To avoid that, we propose working with the full multi-dimensional residual signal. With this, we fully exploit the available analytical redundancy, and the number of rows in $\mathcal{K}_L$ determines the degree of freedom in the design. 

By the block triangular structure of $\mathcal{G}_L$, it is easy to show $\mathrm{dim}\,\mathcal{N}(\mathcal{G}^{\top}_L)=\mathrm{dim}\,\mathcal{N}(\mathcal{O}^{\top}_L)$. Moreover, since $(A,\,C)$ is observable, if $n_\mathrm{o} \leq n$ is the observabilty index of $(A,\,C)$, for any $L>n_\mathrm{o}, \mathcal{O}_L$ has full column rank with row excess of $Ln_y-n$. This gives the following result.
\begin{lem}\label{lem:existential-KL}
	Let $n_\mathrm{o}$ be the observability index of $(A,\,C)$ and Assumption~\eqref{asmp:system-observability} hold. For $L>n_\mathrm{o}$, there exists a full row-rank $\mathcal{K}^y_L\in \mathbb{R}^{(Ln_y-n) \times Ln_y}$ such that $\mathcal{K}^{y}_L\mathcal{O}_L=0$.
\end{lem}

For the sake of brevity, we develop the foundations in this section without considering noise, i.e., $e_k=0$. The effect of noise is empirically investigated in Section \ref{sec:numerical}.

By choosing $L$ according to Lemma~\ref{lem:existential-KL}, the multi-dimensional residual signal at time instant $k$ satisfies
\begin{equation}\label{eq:residual-noisefree}
	r_k = \pi_L\mathbf{f}_{k,L}\,;\quad \pi_L:=\mathcal{K}^{y}_L\mathcal{T}^f_L \in\mathbb{R}^{(Ln_y-n)\times Ln_f}\,.
\end{equation}
This implies that the residual signal belongs to the column space of $\pi_L$ through the linear map $\mathcal{F}: \mathbb{R}^{Ln_f}\to \mathbb{R}^{Ln_y-n}$, where $\mathcal{F}(x):= \pi_Lx$. Thus, the residual signal is a linear combination of fault signals present in the system
\begin{equation*}
	r_k = \sum_{i=1}^{n_f} \pi^{(i)}_L\mathbf{f}^{(i)}_{k,L}\,;\quad \pi^{(i)}_L\in\mathbb{R}^{(Ln_y-n)\times L}\,.
\end{equation*}
where the column vector $\mathbf{f}^{(i)}_{k,L}$ includes only values corresponding to $i$th fault signal over the filter horizon. 
\subsection{Fault dictionaries}
In this section, we show how to construct $\pi^{(i)}_L$ from of I/O data in a data-driven FI framework. First, the fault block-Toeplitz matrix $\mathcal{T}^f_L$ is expressed based on \eqref{eq:fault-matrices-def}. Then, we show how these bases are transformed by pre-multiplying them with $\mathcal{K}^{y}_L$. To do so, the range of $\mathcal{T}^u_L$ is needed to be recovered from the fault-free I/O data.
\begin{asmp}[data rank condition]\label{asmp:PE-signal} The healthy data is recorded such that $\mathrm{rank}{\begin{bmatrix}
				X_{k,L} \\
				\mathbf{U}_{k,L}
		\end{bmatrix}} = n+Ln_u$.
\end{asmp}
\begin{rmk}
		In practice, Assumption~\ref{asmp:PE-signal} can be ensured when the data is collected in open-loop using an input signal $u_k$ that is persistently exciting (PE) of sufficient order \cite{verhaegen1994innovationform}.
	\end{rmk}
\begin{prop}\label{prop:DD-OL-TuL-columnspace}
	Suppose the input signal satisfies Assumption~\ref{asmp:PE-signal}. Given fault-free data $\{u_k,\,y_k\}^N_{k=1}$, we have $\mathcal{R}(\mathcal{T}^u_L)=\mathcal{R}(L_{21})$ and $\mathcal{R}(\mathcal{O}_L)=\mathcal{R}(L_{22})$ where $L_{21}$ and $L_{22}$ are derived from the following LQ decomposition
	\begin{equation*}
		\begin{bmatrix}
			\mathbf{U}_{k,L} \\ \mathbf{Y}_{k,L}
		\end{bmatrix} = \begin{bmatrix}
			L_{11} &0  \\
			L_{21} &L_{22}
		\end{bmatrix}\begin{bmatrix}
			Q^{\top}_1\\Q^{\top}_2
		\end{bmatrix}\,.
	\end{equation*}
\end{prop}
\begin{proof}
	Assumption~\ref{asmp:PE-signal} ensures that the row spaces of $\mathcal{O}_LX_{k,L}$ and $\mathcal{T}^u_L\mathbf{U}_{k,L}$ remain disjoint in the data-equation~\eqref{eq:dataequation-hankel-overwindow}. Thus, we conclude that $L_{21}Q^{\top}_1$ is the term linking $\mathbf{U}_{k,L}$ to $\mathbf{Y}_{k,L}$ because it forms a linear combination of the same bases as in $\mathbf{U}_{k,L}=L_{11}Q^{\top}_1$. Consequently, we obtain $\mathcal{T}^u_L\mathbf{U}_{k,L}=L_{21}Q^{\top}_1$. By substituting $\mathbf{U}_{k,L}$ from the LQ-decomposition, we arrive at $\mathcal{T}^u_LL_{11}Q^{\top}_1=L_{21}Q^{\top}_1$, which implies $\mathcal{T}^u_L=L_{21}L^{-1}_{11}$. The PE input signal guarantees that $L_{11}$ is invertible and full rank, leading to $\mathcal{R}(\mathcal{T}^u_L)=\mathcal{R}(L_{21})$. Moreover, since $\mathcal{O}_LX_{k,L}=L_{22}Q^{\top}_2$, a similar argument establishes that $\mathcal{R}(\mathcal{O}_L)=\mathcal{R}(L_{22})$, thus completing the proof.
\end{proof}

Looking at the definition of fault matrices \eqref{eq:fault-matrices-def}, we observe that the columns of $\mathcal{T}^u_L$ and $I_{n_y}$ serve as the building blocks of $\mathcal{T}^f_L$. That is, $\mathcal{R}(\mathcal{T}^f_L)\subseteq\mathcal{R}(\mathcal{T}^u_L)$ in case of faulty actuators, as well as $\mathcal{R}(\mathcal{T}^f_L)\subseteq\mathcal{R}(I_{L}\otimes I_{n_y})$ for faulty sensors.

Let $\mathcal{T}^{\mathrm{a}_i}_L$ and $\mathcal{T}^{\mathrm{s}_i}_L$ represent the \emph{signatures} of the $i$th actuator fault and sensor fault, respectively, and define them as
\begin{align*}
    &\mathcal{T}^{\mathrm{a}_i}_L=\left(L_{21}\right)_{\Omega_\mathrm{a}}\,;\, \mathcal{T}^{\mathrm{s}_i}_L=\left(I_{L}\otimes I_{n_y}\right)_{\Omega_\mathrm{s}}\,,\\
    &\Omega_\mathrm{a}=\{k\in\mathbb{N}_0\,:\, k=i+t\,n_u,\,t\in\mathbb{N}_0,\,k\leq (L-1)n_u+i\}\,,\\
    &\Omega_\mathrm{s}=\{k\in\mathbb{N}_0\,:\, k=i+t\,n_y,\,t\in\mathbb{N}_0,\,k\leq (L-1)n_y+i\}\,,
\end{align*}
where sets $\Omega_\mathrm{a}$ and $\Omega_\mathrm{s}$ specify the columns to be selected from the corresponding matrices. Then, we introduce \textit{fault dictionaries} as
\begin{equation}\label{eq:faultDict-definition}
	\text{Dictionary}:\begin{cases}
		\mathcal{D}^{\mathrm{a}_i}_L=\mathcal{K}^{y}_L\, \mathcal{T}^{\mathrm{a}_i}_L; 	&\hspace{-5pt} \text{if}\;i\text{th actuator is faulty,} \\
		\mathcal{D}^{\mathrm{s}_i}_L=\mathcal{K}^{y}_L \mathcal{T}^{\mathrm{s}_i}_L;	&\hspace{-5pt} \text{if}\;i\text{th sensor is faulty},
	\end{cases} 
\end{equation} 
in which $\mathcal{D}^{\mathrm{a}_i}_L,\,\mathcal{D}^{\mathrm{s}_i}_L\in\mathbb{R}^{(Ln_y-n)\times L}$.  The total set of dictionaries are collected in $\mathcal{D}_L=\{\mathcal{D}^{\mathrm{a}_i}_L\}^{n_u}_{i=1}\cup \{\mathcal{D}^{\mathrm{s}_i}_L\}^{n_y}_{i=1}$. In addition, denote by $\mathcal{A}:=\{\mathrm{a}_1,\,\dots,\,\mathrm{a}_{n_u}\}$ and $\mathcal{S}:=\{\mathrm{s}_1,\,\dots,\,\mathrm{s}_{n_y}\}$ the sets of actuators and sensors, respectively. Note that learning the dictionaries uses only I/O data, thus not relying on the underlying system matrices.
\subsection{Geometric classifier}
\begin{figure}[tb]
	\centering
	\includegraphics[width=0.7\columnwidth]{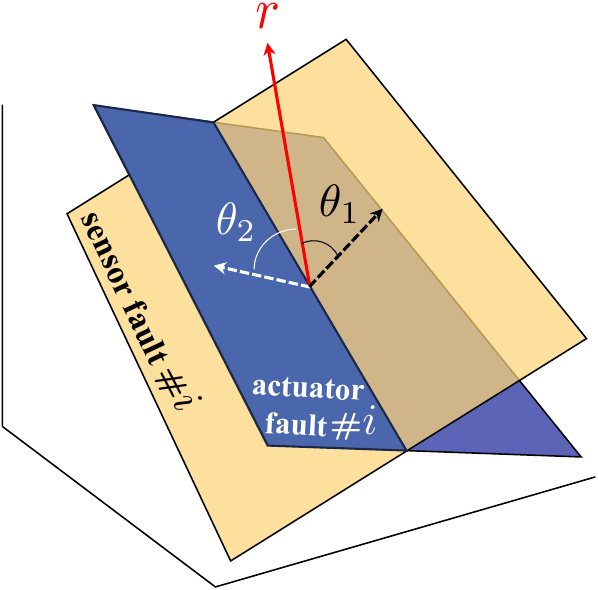}
	\caption{Geometric interpretation of the fault isolation
		problem.}
	\label{fig:hyperplane-intersect}
\end{figure}

In the proposed framework, classification takes place in the subspace $\mathbb{R}^{Ln_y-n}$. In this space, hyperplane $i$ corresponds to the span of the columns in $\mathcal{D}^{(i)}_L\in \mathcal{D}_L$. The residual signal is expected to belong to the subspace that shares the same basis vectors as $\pi_L$; that is, if only one fault is active, the signal should lie entirely within the corresponding subspace for an appropriate choice of $L$. The geometry of the problem suggests that projecting the residual signal onto each hyperplane should provide the necessary information for classification. In other words, the angle between the residual signal and its projection serves as the primary decision variable in the proposed classifier (see Fig~\ref{fig:hyperplane-intersect}). As a result, the problem can be formulated as
\begin{align}\label{eq:cos-theta-classifier}
	\cos{\theta^{(i)}_k}=\frac{\langle r_k,\, \mathbf{P}_{\mathcal{D}^{(i)}_L}[r_k] \rangle}{\lVert r_k\rVert\, \lVert \mathbf{P}_{\mathcal{D}^{(i)}_L}[r_k]\rVert}\,,
\end{align}
with $\mathbf{P}_{\mathcal{D}^{(i)}_L}[r_k]:= \mathcal{D}^{(i)}_L(\mathcal{D}^{(i)^{\top}}_L \mathcal{D}^{(i)}_L)^{-1}\mathcal{D}^{(i)^{\top}}_Lr_k$. The multidimensional time-varying signal $\theta_k$ is defined as $\theta_k :=[\begin{matrix}\theta^{(1)}_k &\cdots &\theta^{(n_u+n_y)}_k\end{matrix}]^{\top}.$ To simplify the exposition, we assume that only \emph{a single fault is present} at time instant $k$ (i.e., $n_f=1$), and the classifier is proposed as
\begin{align*}
	\text{faulty mode}&:= \argmax_{i={1,\,\cdots,\,n_u+n_y}}{\cos{\theta^{(i)}_k}}\,; \quad\quad 0\leq \theta^{(i)}_k\leq \pi/2\,.
\end{align*}
Note that $\cos{x}$ is monotone in the closed interval $[0,\,\frac{\pi}{2}]$.
\begin{rmk}
	The classifier can be extended to handle simultaneous faults, provided the fault subsystem remains left-invertible. This is done by iteratively testing combinations of dictionaries using direct sums of their column spaces, until one yields a zero angle.
\end{rmk}
\section{Fault discernibility}
By introducing the FI scheme, we can determine the conditions under which faults can be perfectly isolated and identify the limiting factors. To emphasize its importance, consider the scenario where the system is subject to only one fault. From Figure~\ref{fig:hyperplane-intersect}, it follows that if the resulting residual signal resides exactly on the intersection of two planes, the angles will be zero for both possible faults. Consequently, the classifier cannot distinguish between them. In the following, we establish the link between fault discernibility and hyperplane intersections based on the filter horizon and the dynamical system properties.

First, we formally define the concept of \emph{mutual discernibility of faults} in terms of the system output behavior.
\begin{dfn}[Mutual discernibility]\label{dfn:mutual-discernibility}
	Let $f^1_k\in\mathbb{R}^{n_1}$ and $f^2_k\in\mathbb{R}^{n_2}$ be two nonzero fault signals with $f^1_k\neq f^2_k$. $f^1_k$ and $f^2_k$ are said to be indiscernible for a given $L$ if there exist $x^1_k,x^2_k$ such that  $\mathbf{y}^1_{k,L}=\mathbf{y}^2_{k,L}$, where $\mathbf{y}^i_{k,L}=\mathcal{O}_Lx^i_k + \mathcal{T}^u_L\mathbf{u}_{k,L} +  \mathcal{T}^{f_i}_L\mathbf{f}^i_{k,L} + \mathcal{T}^e_L\mathbf{e}_{k,L}$. Otherwise, $f^1_k$ and $f^2_k$ are discernible.
\end{dfn}
Next, the null space of $\begin{bmatrix}\mathcal{O}_L & \mathcal{T}^f_L\end{bmatrix}$ is related to the transmission zeros of the fault subsystem, as shown in \cite{sanjeevini2020countingzeros}:
\begin{thm}\label{thm:countingzeros} Let $\tau$ be the smallest integer such that $(A,B_f,C,D_f)$ is $\tau$-delay left invertible, and $\zeta$ be the number of finite and infinite transmission zeros counting multiplicity. Then, if $L \geq \max(\tau,n),$ we have
	\(\mathrm{dim}\,\mathcal{N}(\begin{bmatrix}\mathcal{O}_L & \mathcal{T}^f_L\end{bmatrix}) = \zeta.\)
\end{thm}
\begin{proof}
	Let $\zeta_{\mathrm{fi}}$ be the number of finite transmission zeros counting multiplicity and $\zeta_{\mathrm{inf}}$ be the number of infinite transmission zeros, so that $\zeta = \zeta_{\mathrm{fi}} + \zeta_{\mathrm{inf}}$. By Theorem III.4 of \cite{sanjeevini2020countingzeros}, if $L \geq n$, then 
	\( \mathrm{dim}\,\mathcal{N}(\begin{bmatrix}\mathcal{O}_L & \mathcal{T}^f_L\end{bmatrix}) - \mathrm{dim}\,\mathcal{N}(\mathcal{T}^f_L) = \zeta_{\mathrm{fi}}. \)
	By Theorem IV.8 of \cite{sanjeevini2020countingzeros}, if $L \geq \tau$, $\mathrm{dim}\,\mathcal{N}(\mathcal{T}^f_L) = \zeta_{\mathrm{inf}}.$ Combining the two expressions gives the desired result.
\end{proof}
\begin{lem}\label{lem:nullity-OT-zero-residual}
	For any $\mathbf{f}_0\in\mathbb{C}^{Ln_f}$, $\mathcal{K}^{y}_L\mathcal{T}^f_L\mathbf{f}_0=0$ if and only if there exists $x_0\in\mathbb{C}^n$ such that $\begin{bmatrix}
		\mathcal{O}_L &\mathcal{T}^{f}_L
	\end{bmatrix}\begin{bmatrix}
		x_0 \\ \mathbf{f}_0
	\end{bmatrix}=0$.
\end{lem}
\begin{proof}
	($\Rightarrow$) Suppose $\mathcal{K}^{y}_L\mathcal{T}^f_L\mathbf{f}_0=0$. This implies $\mathcal{T}^{f}_L\mathbf{f}_0\in\mathcal{N}(\mathcal{K}^{y}_L)=\mathcal{R}(\mathcal{O}_L)$. Thus, $\mathcal{T}^{f}_L\mathbf{f}_0=\mathcal{O}_Lv$ for some $v$, and taking $x_0=-v$ yields $\mathcal{O}_Lx_0=-\mathcal{T}^{f}_L\mathbf{f}_0$. ($\Leftarrow$) Left-multiplying the zero equation by $\mathcal{K}^{y}_L$ gives the result.
\end{proof}
A direct consequence of Lemma~\ref{lem:nullity-OT-zero-residual} and Theorem~\ref{thm:countingzeros} is
\begin{equation}
	\mathrm{dim}\,\mathcal{N}(\mathcal{K}^{y}_L\mathcal{T}^f_L)=\mathrm{dim}\,\mathcal{N}(\begin{bmatrix}
		\mathcal{O}_L &\mathcal{T}^{f}_L
	\end{bmatrix})=\zeta\,,
\end{equation}
where $\zeta$ denotes the number of transmission zeros of $(A,B_f,C,D_f)$. Thus, we call a nonzero $\mathbf{f}_0$ satisfying Lemma~\ref{lem:nullity-OT-zero-residual} a \emph{zero-dynamic input}. Moreover, \emph{indiscernible faults} relate to zero-dynamic input directions as follows:
\begin{prop}\label{prop:discernibility-zerodynamicinput}
	Let $\mathbf{f}^1_{k,L} \ne \mathbf{f}^2_{k,L}$ be indiscernible faults. Denote by $\mathbf{f}^{\rm aug}_{k,L}:= [
		f^1_k \;\;\;{-f}^2_k \;\cdots \;f^1_{(L-1)n_1} \;\;{-f}^2_{(L-1)n_2}
	]^\top$ the augmented fault. Then, $\mathbf{f}^{\rm aug}_{k,L}$ is zero-dynamic input of the augmented system $(A,[B^1_f,B^2_f],C,[D^1_f,D^2_f])$. 
\end{prop}
\begin{proof}
	From Definition~\ref{dfn:mutual-discernibility}, indiscernible faults yield identical residuals, i.e., $r^1_{k,L} = r^2_{k,L}$. This implies $\mathcal{K}^{y}_L[\mathcal{T}^{f_1}_L \; \mathcal{T}^{f_2}_L][\mathbf{f}^{1^\top}_{k,L}\;-\mathbf{f}^{2^\top}_{k,L}]^\top=0$. This can be rewritten as $\mathcal{K}^{y}_L\mathcal{T}^{f{\mathrm{aug}}}_L \mathbf{f}^{\mathrm{aug}}_{k,L} = 0$, where $\mathcal{T}^{f{\mathrm{aug}}}_L$ defines the augmented system, obtained by reordering the columns and entries of the associated matrices. By Lemma~\ref{lem:nullity-OT-zero-residual}, $\mathbf{f}^{\mathrm{aug}}_{k,L}$ lies in the zero-dynamic input space of the augmented system.
\end{proof}
\subsection{Intersection of two dictionaries}
Since a fault cannot be isolated when its residual belongs to two different dictionaries, here we characterize the intersection between two dictionaries. In what follows, we present the necessary lemmas and propositions to quantify the intersection in terms of $\mathcal{O}_L$ and $\mathcal{T}^f_L$, enabling the application of the result in Lemma~\ref{lem:nullity-OT-zero-residual}.
\begin{lem}\label{lem:range-intersection}
    Let $P_1\in\mathbb{R}^{m\times n}$ and $P_2\in\mathbb{R}^{m\times n}$. Then, $\mathrm{dim}\left(\mathcal{R}(P_1)\,\cap\,\mathcal{R}(P_2)\right)$ equals to:
    {\renewcommand{\labelenumi}{(\roman{enumi})}
    \begin{enumerate}
	\item $\mathrm{rank}\,P_1 + \mathrm{rank}\,P_2-\mathrm{rank}\,[\begin{matrix}
	P_1   &P_2
	\end{matrix}]$.
	\item $\mathrm{dim}\,\mathcal{N}([\begin{matrix}
	P_1   &P_2
	\end{matrix}])-\mathrm{dim}\,\mathcal{N}(P_1)-\mathrm{dim}\,\mathcal{N}(P_2)$.
	\end{enumerate}
	}
\end{lem}
\begin{proof}
    For the proof of (i), see \cite[Fact 3.14.15]{bernstein2018matrixbook}. Applying the \emph{rank-nullity} theorem to (i) gives (ii).
\end{proof}
\begin{lem}\label{lem:rank-intersec-auxiliary}
    Let $\mathcal{R}({\mathcal{K}^{y}_L}^{\top})=\mathcal{N}(\mathcal{O}^\top_L)$ and $P\in\mathbb{R}^{Ln_y\times L}$. Then, $\mathrm{rank}(\mathcal{K}^{y}_LP)=L-\mathrm{dim}\,\mathcal{N}(\begin{bmatrix}
        \mathcal{O}_L &P
    \end{bmatrix})$.
\end{lem}
\begin{proof}
    By \cite[Proposition~3.6.10]{bernstein2018matrixbook}, $\mathrm{rank}(\mathcal{K}^{y}_LP) = \mathrm{rank}\,P-\mathrm{dim}\left(\mathcal{N}(\mathcal{K}^y_L)\cap \mathcal{R}(P)\right)$. From  the fundamental subspaces theorem, it can be established that $\mathcal{N}(\mathcal{K}^y_L)=\mathcal{R}(\mathcal{O}_L)$. Combining this with the second part of the findings in Lemma~\ref{lem:range-intersection}, we can express the rank as follows
\begin{equation*}
    \mathrm{rank}(\mathcal{K}^{y}_LP) = \mathrm{rank}\,P+\mathrm{dim}\,\mathcal{N}(P)-\mathrm{dim}\,\mathcal{N}(\begin{bmatrix}
        \mathcal{O}_L &P
    \end{bmatrix})\,.
\end{equation*}
    In this derivation, the fact that $\mathrm{dim}\,\mathcal{N}(\mathcal{O}_L)=0$ is already taken into account. Using the \emph{rank-nullity} theorem then yields the desired result.
\end{proof}

The following lemma characterizes the common subspace spanned by the columns of two distinct fault dictionaries. For this, let $c,c'\in\mathcal{A}\cup\mathcal{S}$ and assume that $\mathcal{D}^{c}_L$, $\mathcal{D}^{c'}_L$ with $c \neq c'$ follow \eqref{eq:faultDict-definition}. Define $d^{c,c'}_{\cap}:=\mathrm{dim}\left(\mathcal{R}(\mathcal{D}^{c}_L)\,\cap\,\mathcal{R}(\mathcal{D}^{c'}_L)\right)$.
\begin{lem}\label{lem:dim-intersec}
    The dimension of intersection of subspaces $\mathcal{D}^{c}_L$ and $\mathcal{D}^{c'}_L$ is given as
    \begin{align}\label{eq:dim-intersect-nullspacebased}
        d^{c,c'}_{\cap}&=\mathrm{dim}\,\mathcal{N}(\begin{bmatrix}
        \mathcal{O}_L &\mathcal{T}^{c}_L &\mathcal{T}^{c'}_L
    \end{bmatrix})-\mathrm{dim}\,\mathcal{N}(\begin{bmatrix}
        \mathcal{O}_L &\mathcal{T}^{c}_L
    \end{bmatrix})\notag\\
    &-\mathrm{dim}\,\mathcal{N}(\begin{bmatrix}
        \mathcal{O}_L &\mathcal{T}^{c'}_L
    \end{bmatrix})\,.
    \end{align}
\end{lem}
\begin{proof}
    In light of Lemma~\ref{lem:range-intersection}, we have
\begin{align*}
	d^{c,c'}_{\cap}&=\mathrm{rank}\,\mathcal{D}^{c}_L+\mathrm{rank}\,\mathcal{D}^{c'}_L-\mathrm{rank}\begin{bmatrix}
	    \mathcal{D}^{c}_L &\mathcal{D}^{c'}_L
	\end{bmatrix} \notag\\
    &=\mathrm{rank}(\mathcal{K}^y_L\mathcal{T}^{c}_L)+\mathrm{rank}(\mathcal{K}^y_L\mathcal{T}^{c'}_L)-\mathrm{rank}(\mathcal{K}^y_L[\mathcal{T}^{c}_L\; \mathcal{T}^{c'}_L])\,.
\end{align*}
By applying Lemma~\ref{lem:rank-intersec-auxiliary} to each term individually, the final result is achieved.
\end{proof}
\begin{asmp}\label{asmp:observability-of-subsystems}
    The pair $(A,C_{\mathcal{I}_y})$ is observable for any subset $\mathcal{I}_{y}\subseteq\{1,\,\dots,\,n_y\}$, where $C_{\mathcal{I}_y}$ consists of the rows of $C$ corresponding to the outputs in $\mathcal{I}_{y}$.
\end{asmp}
The main result is presented in the following theorem, where the dimension of the intersection is derived for different combination of fault dictionaries. Each case is defined by $c$ and $c'$, denoting either an $i$th actuator $\mathrm{a}_i$ or an $i$th sensor $\mathrm{s}_i$.
\begin{thm}\label{thm:dim-intersection-zeros}
    Suppose that Assumption~\ref{asmp:leftinvert-faultsubsys} and Assumption~\ref{asmp:observability-of-subsystems}  hold, and let $L\geq n$. Define $\mathcal{I}_{u}\subseteq\{1,\,\dots,\,n_u\}$ and $\mathcal{I}_{y}\subseteq\{1,\,\dots,\,n_y\}$ to be subsets of the inputs and outputs, respectively. Denote by $\zeta^m$, $\zeta^{\mathcal{I}_{u}}$, and $\zeta^{m,\mathcal{I}_{y}}$ the total number of finite and infinite transmission zeros in the subsystem: from the $m$th input to all outputs, from all inputs in $\mathcal{I}_{u}$ to all outputs, and from the $m$th input to all outputs in $\mathcal{I}_{y}$, respectively. Then, for any two different channels $c$ and $c'$
    \begin{align*}
	d^{c,c'}_{\cap}=\begin{cases}
        \zeta^{\{i,j\}}-\zeta^i-\zeta^j &\text{if } c=\mathrm{a}_i, c' = \mathrm{a}_j, i \neq j\,, \\
	\zeta^{i,\{1,\dots,n_y\}\backslash j} &\text{if }          c=\mathrm{a}_i, c' = \mathrm{s}_j\,, \\
        0 &\text{if } c=\mathrm{s}_i, c' = \mathrm{s}_j, i \neq j,n_y>2\,,\\
        n &\text{if } c=\mathrm{s}_i, c' = \mathrm{s}_j, i \neq j,n_y=2\,.
	\end{cases}
    \end{align*}
\end{thm}
\begin{proof}
    To prove this, we primarily rely on the dimension of the intersection \eqref{eq:dim-intersect-nullspacebased} in Lemma~\ref{lem:dim-intersec}. The key aspect is computing the dimension of each nullspace, where Theorem~\ref{thm:countingzeros} connects it to transmission zeros.

    (First case): In this case, fault dictionaries correspond to actuator faults. Direct application of Theorem~\ref{thm:countingzeros} yields $\mathrm{dim}\,\mathcal{N}(\begin{bmatrix}
        \mathcal{O}_L &\mathcal{T}^{\mathrm{a}_i}_L
    \end{bmatrix})=\zeta^i$ and $\mathrm{dim}\,\mathcal{N}(\begin{bmatrix}
        \mathcal{O}_L &\mathcal{T}^{\mathrm{a}_j}_L
    \end{bmatrix})=\zeta^j$. To obtain $\mathrm{dim}\,\mathcal{N}(\begin{bmatrix}
        \mathcal{O}_L &\mathcal{T}^{\mathrm{a}_i}_L &\mathcal{T}^{\mathrm{a}_j}_L
    \end{bmatrix})$, we rearrange the columns so that Theorem~\ref{thm:countingzeros} can be invoked for the subsystem from $\{i,\,j\}$th inputs to all outputs. Note that the rank of a matrix remains invariant under column-wise permutations. Consequently, the nullity for this subsystem is equal to $\zeta^{\{i,j\}}$. Substituting each part into \eqref{eq:dim-intersect-nullspacebased} gives the result.

    (Second case): This case considers an actuator fault versus a sensor fault. We analyze each term separately. According to Theorem~\ref{thm:countingzeros}, $\mathrm{dim}\,\mathcal{N}(\begin{bmatrix}
        \mathcal{O}_L &\mathcal{T}^{\mathrm{a}_i}_L
    \end{bmatrix})=\zeta^i$. On the other hand, Assumption~\ref{asmp:leftinvert-faultsubsys} implies that $\begin{bmatrix}
        \mathcal{O}_L &\mathcal{T}^{\mathrm{s}_j}_L
    \end{bmatrix}$ is full column rank, and therefore $\mathrm{dim}\,\mathcal{N}(\begin{bmatrix}
        \mathcal{O}_L &\mathcal{T}^{\mathrm{s}_j}_L
    \end{bmatrix})=0$. To simplify computation of $\mathrm{dim}\,\mathcal{N}(\begin{bmatrix}
        \mathcal{O}_L &\mathcal{T}^{\mathrm{a}_i}_L &\mathcal{T}^{\mathrm{s}_j}_L
    \end{bmatrix})$, we break it down to a few steps. Since one of the dictionaries corresponds to an actuator fault, we can always find at least $\zeta^i$ nonzero linear combinations of columns in $\begin{bmatrix}
        \mathcal{O}_L &\mathcal{T}^{\mathrm{a}_i}_L &\mathcal{T}^{\mathrm{s}_j}_L
    \end{bmatrix}$ that produce zero based on Theorem~\ref{thm:countingzeros}, implying $\mathrm{dim}\,\mathcal{N}(\begin{bmatrix}
        \mathcal{O}_L &\mathcal{T}^{\mathrm{a}_i}_L &\mathcal{T}^{\mathrm{s}_j}_L
    \end{bmatrix})\geq \zeta^i$. We now show that there exists another set of linear combinations that leads to rank deficiency. Let $j$ represent the $j$th output of the system and $\Omega$ denote the set $\{1,\,\dots,\,n_y\}\backslash j$. Consider $z_0\in\mathbb{C}$ to be a transmission zero for the subsystem from $i$th input to all outputs in $\Omega$.
    For notational clarity, we introduce $\mathcal{O}^{\mathcal{I}_y}_L$ to represent the extended observability matrix for the outputs in $\mathcal{I}_y$, and $\mathcal{T}^{m,\mathcal{I}_y}_L$ to represent the Toeplitz matrix corresponding to subsystem from the $m$th input to the outputs in $\mathcal{I}_y$. Then, according to Theorem~\ref{thm:countingzeros}, there exist pairs $(x^{(\omega)}_0,\,u_0)$ for all $\omega\in\Omega$ such that $\begin{bmatrix}
			\mathcal{O}^{1}_L &\mathcal{T}^{i,1}_L
		\end{bmatrix}\begin{bmatrix}
			x^{(1)}_0 \\ u_0
		\end{bmatrix}=\cdots=\begin{bmatrix}
			\mathcal{O}^{n_y}_L &\mathcal{T}^{i,n_y}_L
		\end{bmatrix}\begin{bmatrix}
			x^{(n_y)}_0 \\ u_0
		\end{bmatrix}=0$ and $\begin{bmatrix}
			\mathcal{O}^{j}_L &\mathcal{T}^{i,j}_L
		\end{bmatrix}\begin{bmatrix}
			x^{(j)}_0 \\ u_0
		\end{bmatrix}\neq 0$. Permuting rows, which does not affect rank, results in
	\begin{align*}
		\mathrm{perm}\,[\begin{matrix}
			\mathcal{O}_L &\mathcal{T}^{\mathrm{a}_i}_L &\mathcal{T}^{\mathrm{s}_j}_L
		\end{matrix}] =
		\left[\begin{array}{ccc} 
			\mathcal{O}^{\Omega}_L &\mathcal{T}^{i,\Omega}_L &0 \\[1pt]
			\hline\\[-1em]
			\mathcal{O}^{j}_L &\mathcal{T}^{i,j}_L &I_L
		\end{array}\right]\,.
	\end{align*}
	Our goal is to demonstrate the existence of a nonzero tuple $\lambda:=(\lambda_1,\lambda_2,\lambda_3)$ satisfying
	\begin{align*}
		\left[\begin{array}{ccc} 
			\mathcal{O}^{\Omega}_L &\mathcal{T}^{i,\Omega}_L &0 \\[1pt]
			\hline\\[-1em]
			\mathcal{O}^{j}_L &\mathcal{T}^{i,j}_L &I_L		\end{array}\right]\begin{bmatrix}
		    \lambda_1\\ \lambda_2 \\  \lambda_3
		\end{bmatrix}=0\,.
	\end{align*}
    Since $z_0$ is also a transmission zero for the subsystem $(i,\Omega)$, it is always possible to find a nonzero $(\lambda_1,\lambda_2)$ such that the top partition becomes zero, i.e., $\mathcal{O}^{\Omega}_L \lambda_1 + \mathcal{T}^{i,\Omega}_L\lambda_2 =0$. Substituting the solution of the top partition into the bottom partition gives $\lambda_3 =-\mathcal{O}^{y_j}_L \lambda_1 - \mathcal{T}^{(u_i,y_j)}_L\lambda_2\neq 0$ as $z_0$ is not a transmission zero for subsystem from the $i$th to the $j$th output; this argument holds for all $\zeta^{i,\Omega}$ transmission zeros. It is easy to show that if there were another output channel not sharing $z_0$ as a transmission zero, the given system of equations would have no nonzero solution, but the proof is omitted for brevity. Assumption~\ref{asmp:leftinvert-faultsubsys} guarantees that no additional linear dependencies exist. Thus, $\mathrm{dim}\,\mathcal{N}(\begin{bmatrix}
        \mathcal{O}_L &\mathcal{T}^{\mathrm{a}_i}_L &\mathcal{T}^{\mathrm{s}_j}_L
    \end{bmatrix})=\zeta^i+\zeta^{i,\Omega}$. The desired result follows from \eqref{eq:dim-intersect-nullspacebased}.
    
    (Third case): In case of two distinct sensor fault dictionaries, $\mathcal{T}^{\mathrm{s}_i}_L$ and $\mathcal{T}^{\mathrm{s}_i}_L$ have independent columns by construction. Moreover, Assumption~\ref{asmp:leftinvert-faultsubsys} ensures that $\begin{bmatrix}
        \mathcal{O}_L &\mathcal{T}^{\mathrm{s}_i}_L
    \end{bmatrix}$ and $\begin{bmatrix}
        \mathcal{O}_L &\mathcal{T}^{\mathrm{s}_j}_L
    \end{bmatrix}$ are full column rank, resulting in a nullspace of zero dimension. Using similar notations as in the second case, but with $\Omega$ being $\{1,\dots,n_y\}\backslash\{i,j\}$, and applying row-wise permutations, we obtain
    \begin{align}\label{eq:permuted-rowwise-sensorVSsensor}
		\mathrm{perm}\,[\begin{matrix}
			\mathcal{O}_L &\mathcal{T}^{\mathrm{s}_i}_L &\mathcal{T}^{\mathrm{s}_j}_L
		\end{matrix}] =\left[\begin{array}{ccc} 
			\mathcal{O}^{\Omega}_L &0 &0 \\
			\hline\\[-1em]
			\mathcal{O}^{i}_L &I_L &0\\
                \mathcal{O}^{j}_L &0 &I_L\\
            \end{array}\right]\,.
    \end{align}
    Assumption~\ref{asmp:observability-of-subsystems} implies that $\mathrm{dim}\,\mathcal{N}(\begin{bmatrix}
        \mathcal{O}_L &\mathcal{T}^{\mathrm{s}_i}_L &\mathcal{T}^{\mathrm{s}_j}_L
    \end{bmatrix})=0$ if $n_y>2$ based on the structure in \eqref{eq:permuted-rowwise-sensorVSsensor}. In case of $n_y=2$, it follows from \eqref{eq:permuted-rowwise-sensorVSsensor} that $\mathrm{dim}\,\mathcal{N}(\begin{bmatrix}
        \mathcal{O}_L &\mathcal{T}^{\mathrm{s}_i}_L &\mathcal{T}^{\mathrm{s}_j}_L
    \end{bmatrix})$ reduces to $\mathrm{dim}\,\mathcal{N}(\begin{bmatrix}
        \mathcal{O}^i_L \\ \mathcal{O}^j_L
    \end{bmatrix})=n$. This concludes the proof.
\end{proof}
\begin{rmk}
    When the underlying system has only two outputs ($n_y=2$), the dimension of intersections in all cases of Theorem~\ref{thm:dim-intersection-zeros} coincides with the system order, i.e., for any two distinct channels $c,c'\in\mathcal{A}\cup\mathcal{S}$, we have $d^{c,c'}_\cap=n$.  
\end{rmk}
\begin{figure}[tb]
	\centering	\includegraphics[width=\columnwidth]{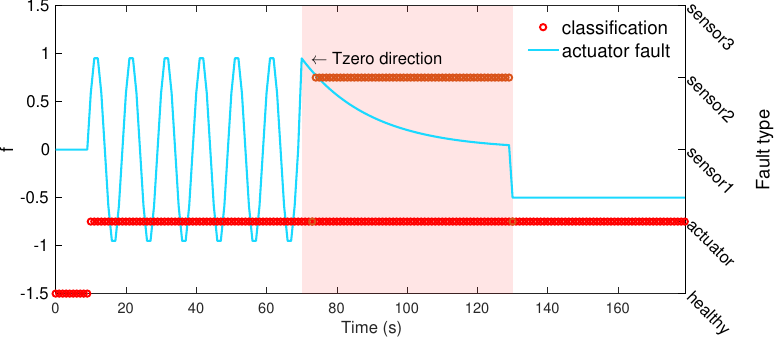}
	\vspace{-1.5em}
	\caption{A noise-free actuator fault scenario. The fault magnitude is indicated in the left y-axis, while the classification result follows the right y-axis.}
	\label{fig:FI-performance-noisefree}
\end{figure}
\begin{figure*}[tb!]
	\centering
	\includegraphics[width=0.6\textwidth]{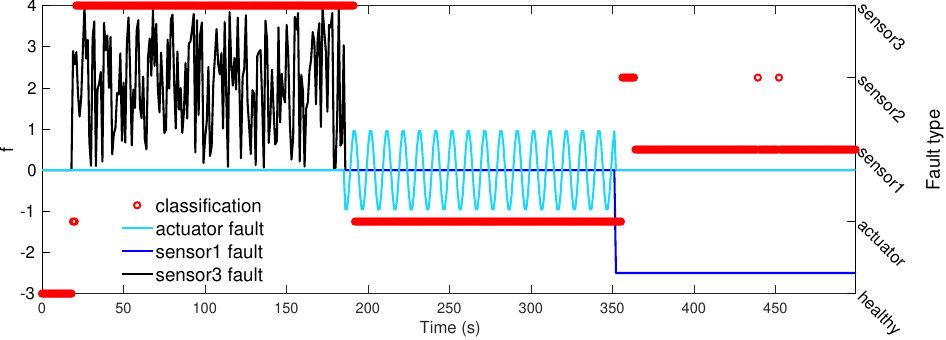}\hspace{10pt}
	\includegraphics[width=0.5\columnwidth]{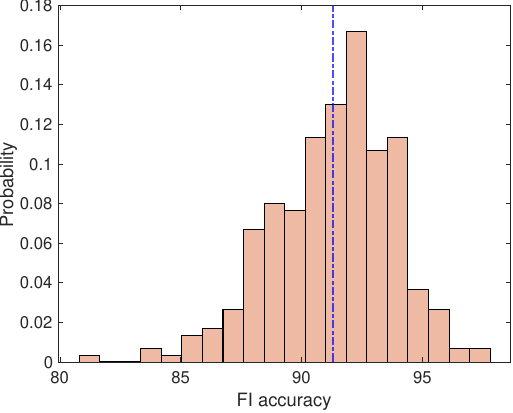}
	\caption{The proposed FI performance in the presence of noise with different types of faults. (Left) The left y-axis represents fault signals in different components, while the classification result is depicted with respect to the right y-axis. (Right) The corresponding histogram.}
	\label{fig:FI-performance-noisy}
\end{figure*}
\section{NUMERICAL EXAMPLES}\label{sec:numerical}
To showcase the theoretical results, we design experiments under both noise-free and noise-contaminated conditions. Consider the following LTI system
\begin{align*}
		A&=\begin{bmatrix}
			0 &1 &0 &0 \\
			0 &0 &1 &0 \\
			0 &0 &0 &1 \\
			-0.136 &0.956 &-2.406 &2.580
		\end{bmatrix},\; B_u=\begin{bmatrix}
			2.520\\3.147\\2.945\\2.458
		\end{bmatrix} \\
		C&=\begin{bmatrix}
			1 &0 &0 &0 \\
			-0.027 &0.083 &-0.038 &-0.030\\
			0.194 &-0.868 &1.234 &-0.566
		\end{bmatrix},\; D_u=\begin{bmatrix}
			1\\1\\1
		\end{bmatrix}\,.
\end{align*}
In this system of order $n=4$, only the channels from the input to the outputs $\{1,3\}$ share a transmission zero at $z=0.95$, i.e., $\zeta^{1,\{1,3\}}=1$.

\textbf{Scenario~1.} In a noise-free setting, the considered system is subject to only an actuator fault throughout the simulation. To highlight some of the key findings in Theorem~\ref{thm:dim-intersection-zeros},  it is assumed that the fault dictionaries are computed with nominal values. The input signal is a multi-step signal taking values from $\{1,2,1.5\}$. We aim to design a fault signal that leads to an isolation confusion. To this end, consider the fault signal as $f_k=\{0\;\text{if}\;k<10,\, \sin{(0.2\pi k)}\;\text{if}\;10\leq k \leq 70,\,(0.95)^{k-70}\;\text{if}\;70\leq k \leq  130,\,-0.5\;\text{otherwise}\}$. The isolation performance is illustrated in Fig.~\ref{fig:FI-performance-noisefree} for $L=5$. From the figure, it can be seen that the classifier cannot distinguish between the actuator and sensor $\#2$ faults during the shaded period (i.e., $\cos{\theta}=1$ for both). This is because the residual signal exactly resides at the intersection of the corresponding hyperplanes in this period. This non-trivial intersection corresponds to the second case of Theorem~\ref{thm:dim-intersection-zeros}, where an actuator and a sensor fault become indiscernible if the subsystem from the actuator to the considered sensor does not contain the transmission zero shared by the other output channels.

\textbf{Scenario~2.} In a more realistic setting, the faulty subsystem switches between multiple fault modes under a noisy condition. The perturbed healthy I/O data is collected using a PRBS signal, generating $N=1000$ samples. The data-generating system is perturbed by both process and measurement noise via the innovation signal, modeled as zero-mean white noise with the following matrices
\begin{equation*}
	\resizebox{\columnwidth}{!}{$\Sigma_e =\begin{bmatrix}
			5.25 &4.73 &3.96\\
			4.73 &4.87 &3.68\\
			3.96 &3.68 &3.59
		\end{bmatrix},\, 
		K=\begin{bmatrix}
			0.1760 &0.6259 &0.0686 \\
			-0.0815 &0.4654 &-0.2711\\
			-0.288 &0.2886 &-0.1961\\
			-0.3268 &0.1314 &0.3146
		\end{bmatrix}\,,$}
\end{equation*}
and the signal-to-noise ratio (SNR) is $25$. Setting $L=15$ to estimate fault dictionaries and to obtain $\mathcal{K}_L$ by examining the singular values of the LQ decomposition, as outlined in Proposition~\ref{prop:DD-OL-TuL-columnspace}. The FI performance is evaluated through a Monte Carlo simulation. The average FI accuracy across simulations is $91.28\%$. The result for one particular realization is shown in Figure~\ref{fig:FI-performance-noisy}. Apart from some transient behavior, the results demonstrate the effectiveness of the proposed method.
\section{CONCLUSIONS}
The proposed FI approach can perfectly isolate faults under the noise-free condition in case of discernible faults. While the presence of noise may degrade classification performance, the filter still performs satisfactorily. A formal investigation of filter performance under noisy conditions remains an interesting direction for future work.

%%%%%%%%%%%%%%%%%%%%%%%%%%%%%%%%%%%%%%%%%%%%%%%%%%%%%%%%%%%%%%%%%%%%%%%%%%%%%%%%

%\section*{APPENDIX}

%\section*{ACKNOWLEDGMENT}

%%%%%%%%%%%%%%%%%%%%%%%%%%%%%%%%%%%%%%%%%%%%%%%%%%%%%%%%%%%%%%%%%%%%%%%%%%%%%%%%

%==========================================
% USE "txs:///bibtex" backend to compile the bibliography 
%==========================================

\bibliographystyle{IEEEtran}
\bibliography{mybibfile}

\end{document}